\documentclass[twocolumn,aps,floatfix]{revtex4}
\usepackage{amssymb}
\usepackage{graphicx}

\begin{document}

\title{Formation of soliton trains in Bose-Einstein condensates by
       temporal Talbot effect}

\author{Krzysztof Gawryluk,$\,^1$ Miros{\l}aw Brewczyk,$\,^1$
        Mariusz Gajda,$\,^2$ and Jan Mostowski$\,^2$}

\affiliation{\mbox{$^1$ Instytut Fizyki Teoretycznej, Uniwersytet w Bia{\l}ymstoku,
                        ulica Lipowa 41, 15-424 Bia{\l}ystok, Poland}  \\
\mbox{$^2$Instytut Fizyki  PAN, Aleja Lotnik\'ow 32/46,
                        02-668 Warsaw, Poland}  }

\date{\today}

\begin{abstract}
We study the recent observation of formation of matter-wave soliton
trains in Bose-Einstein condensates. We emphasize the role of the
box-like confinement of the Bose-Einstein condensate and find that
there exist time intervals for the opening the box that support the
generation of real solitons.  When the box-like potential is switched
off outside the existing time windows, the number of peaks in a train 
changes resembling missing solitons observed in the experiment. Our 
findings indicate that a new way of generating soliton trains in 
condensates through the temporal, matter-wave Talbot effect is
possible.

\end{abstract}

\maketitle

The notion of soliton belongs to the most popular concepts in physics.
Any localized and long-living structure is readily called a soliton.
Solitons are identified with solutions of nonlinear equations and
appear in many contexts of science ranging from physics (nonlinear
optics, hydrodynamics, particle physics, etc.) to molecular biology.
Solitons are formed because of the existence of nonlinear interaction
in the system which cancels the dispersion and hence allows for the
propagation of shape preserving objects.

In the case of dilute atomic quantum gases the nonlinearity is
determined by the effective interaction between atoms that can be both
repulsive or attractive. For repulsive uniform condensates the
appropriate nonlinear equation (i.e. the Gross-Pitaevskii equation)
predicts dark soliton (a hole in the density associated with a phase
jump) as a solution \cite{Zakharov}. Such excitations of Bose-Einstein
condensates have been already observed in experiments with trapped
alkali atoms \cite{dark}.

Generating bright solitons in atomic quantum gases is more difficult
task because it requires working with attractive condensates. Due to a
collapse, in such samples the number of atoms is limited and small
(see Ref.  \cite{Stringari}). This obstacle has been overcome in
two ways. In the first one, a large repulsive condensate is formed and
then the interactions are changed from repulsive to attractive by
using the Feshbach resonance technique \cite{Hulet, Salomon}, whereas
in the second attempt \cite{lattice} the optical lattice and the notion of
negative effective mass are utilized.

The goal of this Letter is twofold. First, we focus on the experiment of 
Ref. \cite{Hulet} where the trains of bright solitons are generated in 
attractive condensate of $^7$Li atoms. Existing theoretical explanations 
involve quantum phase fluctuations \cite{Hulet1} or modulational instability 
\cite{Italians, Carr} as the main mechanism leading to the observed structures. 
None of these papers, however, attempt to model the experiment of Ref. 
\cite{Hulet}. In our Letter we report on calculations that correspond to the
same values of all parameters as in Ref. \cite{Hulet} with realistic
three-dimensional geometry. Secondly, we concentrate on the role of the 
box-like confinement and show that there exists a regime where multiple 
scattering of gas from the walls of the box might result in formation of 
``real'' solitons, i.e. multipeak structures that undergo elastic collisions. 
We suggest a new version of the experiment and specify physical conditions
supporting the formation of ``genuine'' soliton trains.
The underlying physics of our proposition can be understood with the 
help of the temporal Talbot effect, already experimentally observed 
in the context of Bose-Einstein condensate of sodium atoms diffracted 
by the pulsed grating \cite{Phillips}.

To describe the experiment of Ref. \cite{Hulet} we start, as the Ref. 
\cite{Italians} does, with the time-dependent dissipative Gross-Pitaevskii 
equation
\begin{equation}
i\hbar \frac{\partial \psi}{\partial t} = \left(-\frac{\hbar^2}{2m} \nabla^2
+V_{tr} + g|\psi|^2 - i \gamma |\psi|^4 \right) \psi  \;,
\label{GP}
\end{equation}
where $\psi({\bf r},t)$ is the macroscopic wave function of
Bose-Einstein condensate of $^7$Li atoms,
$V_{tr}(z,\rho)=m\,(\omega_z^2 z^2 + \omega_{\perp}^2 \rho^2)\,/2 +
V_{box}$ is the axially symmetric trapping potential, $g=4\pi \hbar^2
a_s/m$ (with $a_s$ being the scattering length which determines the
strength of the interaction). The number of atoms equals $N=3\times
10^5$, the geometry of the harmonic confinement is given by
$\omega_z=2\pi \times 4$\,Hz and $\omega_{\perp}=2\pi \times 800$\,Hz,
and the box-like potential $V_{box}$ (the ``end caps'') of length of
$L=12$ osc. units is positioned on the slope of the harmonic
potential, shifted by $15$ osc. units from its center. The initial
value of the scattering length of $^7$Li condensate prepared in the
end caps, which is positive and equals $200$\,a$_0$ (a$_0$ -- Bohr
radius), is changed within approximately $10$\,ms to its final value
$a_s=-3$a$_0$ and then the system is kept in the box-like potential
for further $17$\,ms. Finally, the end caps are turned off. As opposed
to the Ref. \cite{Italians}, our calculational parameters are the same 
as in the experiment of Ref. \cite{Hulet}.

\begin{figure*}[thb]
\resizebox{7.0in}{1.in}
{\includegraphics{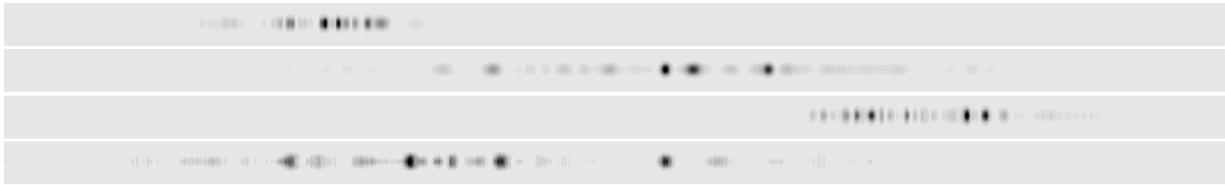}}
\caption{Moving soliton train of $^7$Li atoms near the turning points
and at the center of the oscillation. Successive images correspond to
6 ms, 48 ms, 125 ms, and 185 ms after the end caps are off. The number
of atoms, the trap parameters, and the way the scattering length is
changed are the same as in the experiment of Ref. \cite{Hulet}. Color
images are available at \cite{color}. }
\label{train}
\end{figure*}

The imaginary term in Eq. (\ref{GP}) describes the losses due to
three-body recombination processes \cite{Kagan}, \cite{Italians}. 
Its presence is necessary to get the agreement with experimental
results. Since the losses were not investigated experimentally, we
follow the references just mentioned and put $\gamma=2.05\times
10^{-26}$\,cm$^6$\,s$^{-1}$.  The dissipative term is turned on,
according to the observation in \cite{Hulet}, when the interaction
strength becomes negative and is strongly diminished (by a factor of
$100$) when the end caps are off. When the losses are kept constant
during the whole calculations the final number of atoms is much less
than the number observed in the experiment. On the other hand, 
decreasing the dissipative term leads to the condensate collapse. Another 
way of getting agreement with experimental data (i.e. the final number of 
atoms) is to lower the initial number of atoms while keeping the same value 
of $\gamma$ all the time. Hence, the dissipative term plays a crucial role 
in stabilizing the system. Certainly, it has an influence on the collapse 
studied recently in Ref. \cite{Carr}.

After switching off the box potential the bosonic cloud starts to
oscillate in harmonic trap. As it is illustrated in Fig. \ref{train},
the clouds actually breaks into several peaks which propagate in the
potential for many oscillatory cycles. However, we observe that
certain peaks disappear during the evolution and reappear again
later. For example, in the second frame of Fig. \ref{train} when the
condensate goes through the oscillatory center from the left to the
right, the number of distinguishable peaks is $3$, whereas in the
lowest frame (the system goes now to the left) that number is
increased to $4$. Such structures can not be, in fact, considered as
solitons.

\begin{figure}[thb]
\resizebox{3.0in}{2.2in}
{\includegraphics{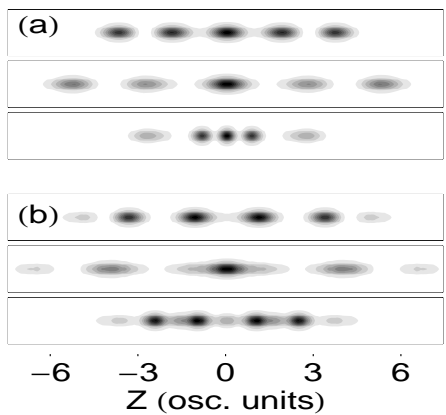}}
\caption{Illustration of the importance of a time when the end caps
are switched off. Frame (a) shows the condensate density at $30$ ms,
$66$ ms, and $108$ ms after the box is off, i.e. $61$ ms after
the interaction strength is changed. This is the case of the time
window when, as a result, the structure consisted of the same number
of peaks is formed. As opposed, frame (b) resembles
Fig. \ref{train}. Here, the number of peaks changes during the
evolution. Successive snapshots correspond to $30$ ms, $66$ ms, and
$139$ ms after the end caps are turned off outside the time window
($52$ ms after the change of the interaction strength). }
\label{density}
\end{figure}

Explaining the experimental results of Ref. \cite{Hulet} we, as
opposed to other theoretical works \cite{Hulet1, Italians, Carr},
emphasize the role of the box-like potential. We demonstrate how 
important is the time when it is open and its location. To this end, 
we have performed calculations for a smaller sample of $N=10^4$ $^7$Li 
atoms created with initial scattering length $a_s=100$a$_0$, which is 
next changed within $10$\,ms to the final value $a_s=-3$a$_0$. The 
dissipative term is kept constant all the time. Our results are 
summarized in Fig. \ref{density}.  The end caps are positioned 
symmetrically with respect to the center of the harmonic trap. Only 
in such a configuration we discover the existence of time windows, i.e. 
the time intervals for the opening the end caps that support the 
generation of train of real solitons (multipeak structures with 
preserved number of peaks).  This is illustrated in Fig. \ref{density}, 
where the size of the box-like potential is $L=4.0$ osc. units 
$(75.6\, \mu {\rm m})$. Frame (a) shows the density in the case when 
the end caps are switched off within the time window ($61$ ms after the 
interaction strength is changed). Here the number of peaks is still the 
same.  On the contrary, when the end caps are off at the time outside 
the time window, the number of peaks changes resembling the missing 
solitons observed in the experiment \cite{Hulet}. Calculations also 
show the existence of further time windows, especially for shorter end 
caps. The duration of the time window is about $3$\,ms and increases when 
the size of the end caps gets larger.

\begin{figure}[thb]
\resizebox{3.in}{3.0in}
{\includegraphics{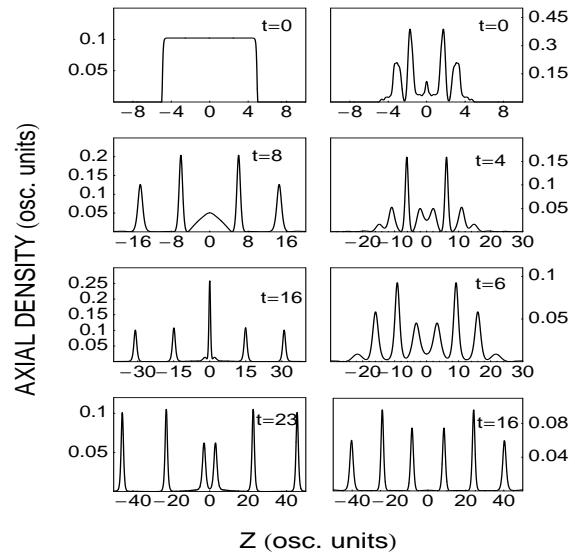}}
\caption{Axial density profiles of a condensate of $10^4$ $^7$Li atoms
during the evolution under no axial confinement. The size of the end caps 
equals $189\, \mu {\rm m}$. The left column corresponds to Fig. 1 in 
Ref. \cite{Italians} when time window condition is not fulfilled. For the 
right column the end caps are turned off within the time window ($314$\,ms 
after the scattering length is changed). In both cases the time $t=0$
means the time when the end caps are removed.}
\label{Itafig}
\end{figure}

It is important to understand the role of the location of the box-like
potential. We have checked numerically that the time windows persist even 
if the end caps are shifted from the center of the harmonic potential by 
several percent, depending on the size of the end caps. Moreover, we do not 
see the time windows when the size of the end caps is too large. It is clear 
that the discussed phenomenon is somehow spoiled by the presence of the 
harmonic trap. As will be explained later, the disappearance of time 
windows is caused by the loss of Talbot-type recurrence due to nonuniform
shift of energy levels of the box potential forced by the harmonic trap.
Certainly, the most favorable conditions for the existence 
of time windows are those when there is no axial harmonic confinement. Such 
calculations have been already performed, see Fig. 1 in Ref. \cite{Italians}. 
However, the authors of Ref. \cite{Italians} have not investigated carefully 
the influence of the delay time between the removal of the end caps and the 
switching off the scattering length and they overlooked the possibility of 
existence of time windows. In Fig. \ref{Itafig} we plot the axial density 
profiles in the case when the box-like potential is turned off just at the 
time when the scattering length is changed (left column) and in the case when 
the box is removed within the time window. The left column shows that, in fact, 
the condensate is split into $6$ peaks (not $5$ as reported in \cite{Italians})
and its evolution is violent in a sense that the number of peaks changes for 
a long duration. On the other hand, as demonstrated in the right column, the 
number of peaks (solitons) is settled much earlier when the end caps are 
switched off within the time window.

The origin of time windows is due to the temporal Talbot effect. Therefore,
we start with a purely linear case and consider the evolution of the 
symmetric wave packet in an ``open'' rectangular one-dimensional box. The 
walls of the box are very high and therefore we will expand the initial 
wave packet in the basis of the symmetric eigenstates of infinitely deep 
well potential that are $\sqrt{2/L} \cos{(k_n x)}$ with 
$k_n=\frac{\pi}{L} (2 n +1)$, where $n=0,1,...$ and $L$ is the length of 
the well. Note that the wave functions are zero in the classically forbidden 
region. The time evolution reads
\begin{equation}
\psi(x,t) = \sum_n \alpha_n e^{-i E_0 (2n+1)^2 t/\hbar} \,
\sqrt{\frac{2}{L}} \cos{k_n x}  \,,
\label{exp}
\end{equation}
where $E_0=\hbar^2 \pi^2/(2mL^2)$ and $\{\alpha_n\}$ are the coefficients of
the expansion of $\psi(x,0)$.
Fourier transform of (\ref{exp}) is given by
\begin{eqnarray}
\tilde{\psi}(k,t) & = & \sqrt{\frac{L}{2}} \sum_n \alpha_n
e^{-i E_0 (2n+1)^2 t/\hbar}  \nonumber  \\
& \times &\left( \frac{\sin{\frac{L}{2}(k-k_n)}}{\frac{L}{2}(k-k_n)} +
\frac{\sin{\frac{L}{2}(k+k_n)}}{\frac{L}{2}(k+k_n)} \right) \,.
\label{FT}
\end{eqnarray}
According to the above formula the momentum distribution is recovered
after a period of $T_{rev}=\pi/4$ (in units of $\hbar/E_0$) and this
is the result of the Talbot-type recurrence occurring in the phase
factor ( ``$n(n+1)$'' dependence on the quantum number $n$).
However, this formula exhibits more structure. At time
$T_{win}^{lin}=\pi/8$ the resulting momentum distribution forms a set
of fully separated groups centered at momenta $k=k_n$. This can be
verified (e.g. numerically) taking into account several facts: (1) the
form of the phase factor in Eq. (\ref{FT}); (2) the localization of
the function $\sin(x)/x$ around $x=0$; (3) assumed weak dependence of
coefficients $\alpha_n$ on $n$.

If the box-like potential is switched off at times around
$T_{win}^{lin}=\pi/8$ all groups of momenta transform into separated
wave packets which just spread out during the further evolution. This
spreading can be stopped by turning on the nonlinear term and then the
wave packets are transformed into Zakharov solitons \cite{Zakharov}.
Hence, the nonlinearity is essential to built the solitons from the
separated wave packets. Here, the separated wave packets originate
from the temporal Talbot effect combined with the appropriate initial 
conditions. In a real experiment initial conditions are formed as a result 
of modulational instability driven by the condensate collapse.

Looking at the formula (\ref{FT}), it is clear that the presence of the 
box-like potential is essential since it allows the condensate wave function 
for multiple reflexions from the walls resulting in the interference pattern 
found above. However, ``small addition'' of the harmonic potential does not
spoil the picture. This can be proved by the perturbation calculus. It turns
out that for not too large size of the box $L$ ($L<5$ osc. units) all the
levels but the lowest one are shifted approximately uniformly. In such a
way the new time scale that is proportional to the reciprocal of the
difference of the shifts of the two lowest levels (and approximately equals 
the trap period) can be introduced. Based on the formula (\ref{FT}) only those 
time windows survive that occur at times shorter than the trap period (although 
``revivals'' are observed for long enough times).

We can also prove that the temporal Talbot effect survives under the
presence of the weak nonlinearity. For small enough attraction the nonlinear
term that appears in the Gross-Pitaevskii equation can be treated as a
perturbation to the linear case. It turns out that all single-particle levels 
are shifted up uniformly. Therefore, from the formula (\ref{FT}), one should 
expect formation of groups in momentum space at the same times and of the same 
duration as in the linear case.

\begin{figure}[thb]
\resizebox{3.in}{2.2in}
{\includegraphics{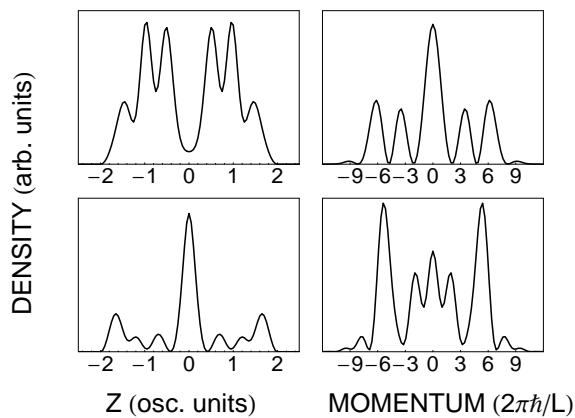}}
\caption{Spatial (left column) and momentum (right column) densities
at the time when the end caps are off. The upper frames illustrate the
case of the time window in which the momentum density shows well
developed groups of momenta. These groups are next transformed to
groups in a position space. The lower frames correspond to the case when
there are no separated peaks in a momentum space and consequently the
probability flow leads to ``missing solitons'' structures.}
\label{main}
\end{figure}

The temporal Talbot effect persists also when the nonlinearity is
larger and comparable to that one present in the experiment of Ref.
\cite{Hulet}. As it is the case, we plot in Fig. \ref{main} the axial 
momentum densities (at the time when the end caps are off) corresponding 
to the frames (a) and (b) of Fig. \ref{density}. Fig. \ref{main} proves 
that the time windows survive under the conditions  when both the box-like 
and harmonic confinements and the nonlinearity are present. When the end 
caps are off just within the time window (the upper frames) the momentum
density shows five distinguishable peaks whereas the density in the
position space consists of two broad peaks with partially
developed three subpeaks. Upper frames in Fig. \ref{density} confirm
that afterward five (not two nor six) solitons are developed. It means
that all momenta groups have been transformed to peaks in position
space. These peaks oscillate with the trap period, collide when they
meet at the center of the trap and then reappear. Their motion is
particle-like. If one considers a point-like particle moving according to
the Newton equation, initially placed at the trap center and
assign the initial velocity determined by the maximum value of the momenta
group, the particle will follow the density peak. Hence, it is
reasonable to use the name solitons for the density peaks.

When the end caps are switched off outside the time window (lower
frames in Fig. \ref{main}) the number of density peaks changes during
the evolution (see the lower part of Fig. \ref{density}). This is
because the momentum density does not consist of well separated groups
and the probability can flow from one group to the other forming the
``missing solitons'' structures.  This is, in fact, the case of
numerical calculations reported in Ref.  \cite{Italians}. Fig. 5 of
Ref. \cite{Italians} shows that the number of peaks changes in time
while the condensate is axially confined. On the other hand, when the
axial confinement is off the condensate always ends in a state with a
fixed number of solitons.

The time the first time window appears is slightly larger than 
$T_{win}^{lin}=\pi/8$ predicted from the linear theory. We found
that it is shifted by an amount of the order of characteristic time 
scale $\hbar/(gN/V)$, which in our case is equal to several milliseconds.

In conclusion, we have shown the importance of the box-like potential in 
experiments like that of Ref. \cite{Hulet}. Contrary to missing solitons 
structures genuine bright solitons can be generated in a repulsive condensate 
by changing the sign of the scattering length with the help of the Feshbach 
resonance technique and keeping the condensate in the end caps for long enough 
time. In fact, what we propose to create solitons is to utilize the Talbot 
effect that is well known from the linear physics. It turns out, however, that 
this effect survives under the condition of weak nonlinearity what was proved by 
using the perturbative analysis but also under the conditions of the experiment 
of Ref. \cite{Hulet} as was demonstrated by direct solving of the Gross-Pitaevskii 
equation. Perhaps, it is not surprising since the temporal Talbot effect was 
already experimentally observed for a condensate of sodium atoms, although in
a different situation when the condensate was diffracted by a pair of pulsed 
gratings and the periodicity with respect to the time delay between pulses
in a series of condensate images was discovered \cite{Phillips}.

\acknowledgments We acknowledge support by the Polish KBN Grant No. 2
P03B 052 24.  Some of our results have been obtained using computers
at the Interdisciplinary Centre for Mathematical and Computational
Modelling of Warsaw University.

\end{document}